\documentclass[pdflatex,sn-mathphys]{sn-jnl}% Math and Physical Sciences Reference Style
\usepackage[T1]{fontenc}
\usepackage{graphicx}
\usepackage{soul}
\usepackage{subfig}
\usepackage{amsmath}
\usepackage{cleveref}
\usepackage{booktabs}
\usepackage{color,soul}
\usepackage{multirow}
\usepackage{makecell}
\usepackage[title]{appendix}

\algnewcommand\algorithmicforeach{\textbf{for each}}
\algdef{S}[FOR]{ForEach}[1]{\algorithmicforeach\ #1\ \algorithmicdo}

\jyear{2022}%

\theoremstyle{thmstyleone}%
\theoremstyle{thmstyletwo}%

\theoremstyle{thmstylethree}%
\newtheorem{definition}{Definition}%

\begin{document}

\title[Mining higher-order motifs in large hypergraphs]{Exact and sampling methods for mining higher-order motifs in large hypergraphs}


\author*[1]{\fnm{Quintino Francesco} \sur{Lotito}}\email{quintino.lotito@unitn.it}

\author[2]{\fnm{Federico} \sur{Musciotto}}\email{federico.musciotto@unipa.it}

\author[3]{\fnm{Federico} \sur{Battiston}}\email{battistonf@ceu.edu}

\author[1]{\fnm{Alberto} \sur{Montresor}}\email{alberto.montresor@unitn.it}

\affil*[1]{\orgdiv{Department of Information Engineering and Computer Science}, \orgname{University of Trento}, \orgaddress{\street{via Sommarive 9}, \city{Trento}, \postcode{I-38123}, \country{Italy}}}

\affil[2]{\orgdiv{Dipartimento di Fisica e Chimica Emilio Segr\`e}, \orgname{Universit\`a di Palermo}, \orgaddress{\street{Viale delle Scienze}, \city{Palermo}, \postcode{I-90128}, \country{Italy}}}

\affil[3]{\orgdiv{Department of Network and Data Science}, \orgname{Central European University}, \orgaddress{\street{Quellenstraße 51}, \city{Vienna}, \postcode{A-1100}, \country{Austria}}}

\abstract{
Network motifs are recurrent, small-scale patterns of interactions observed frequently in a system.
They shed light on the interplay between the topology and the dynamics of complex networks across various domains. 
In this work, we focus on the problem of counting occurrences of small sub-hypergraph patterns in very large hypergraphs, where higher-order interactions connect arbitrary numbers of system units. 
We show how directly exploiting higher-order structures speeds up the counting process compared to traditional data mining techniques for exact motif discovery. 
Moreover, with hyperedge sampling, performance is further improved at the cost of small errors in the estimation of motif frequency.
We evaluate our method on several real-world datasets describing face-to-face interactions, co-authorship and human communication. We show that our approximated algorithm allows us to extract higher-order motifs faster and on a larger scale, beyond the computational limits of an exact approach.}

\keywords{Network motifs, Hypergraph algorithms, Higher-order networks, Complex networks}

\maketitle

\section{Introduction}
Network motifs are recurring patterns of interactions among a small set of nodes that appear in an observed network at a significant frequency.
Motif analysis has established itself as an important tool for investigating networks at their microscale, highlighting the interdependence between the topology and dynamics of real-world networked systems~\cite{milo2002network,schwarze2020motifs}. In fact, interacting systems with close functionalities tend to display similar over- and under-represented patterns of interactions~\cite{milo2004superfamilies}.

Network motifs have found a vast set of applications in a number of different domains, such as biology~\cite{alon2007network,shen2002network,dobrin2004aggregation}, neuroscience~\cite{sporns2004motifs}, medicine~\cite{chen2013identification}, social network analysis~\cite{hong2014social}, finance~\cite{saracco2016detecting} and ecology~\cite{bascompte2009assembly,simmons2019motifs}.

Given their multiple real-world applications, it is not surprising that the notion of network motifs has been extended to a variety of richer and more flexible network models, including weighted~\cite{onnela2005intensity}, temporal~\cite{kovanen2011temporal,paranjape2017motifs} and multilayer~\cite{battiston2017multilayer,kivela2018isomorphism} networks.
Recently, growing interest has been devoted to modelling real-world systems with group interactions~\cite{battiston2020networks,battiston2021physics}, from co-authorship~\cite{patania2017shape} to face-to-face~\cite{cencetti2021temporal} interactions, by exploiting more complex mathematical tools such as hypergraphs~\cite{berge1973graphs}, where hyperedges encode relationship among an arbitrary number of units.
In hypergraphs, it is possible to identify higher-order motifs, i.e., connected sets of nodes interacting not only through pairwise edges but also through hyperedges encoding association among three or more nodes.
In particular, higher-order motifs can be defined in terms of the overlapping patterns of hyperedges of a fixed size~\cite{lee2020hypergraph}, or, more traditionally, by investigating all possible patterns of connected sub-hypergraphs for a given number of nodes~\cite{lotito2021higher}.

Independently on the underlying network model, all the algorithms for motif analysis involve the following steps: (i) counting the occurrences of each motif in an observed network, (ii) counting the occurrences of each motif in suitable randomizations of the observed network and (iii) evaluating the over- and under-expression of each motif.
The problem of counting the frequency of each motif in a target large network is inherently computationally challenging since it is equivalent to the problem of subgraph isomorphism, a well-known NP-complete problem.
Moreover, for the statistical evaluation of motif frequencies, the counting step is repeated on each sample of the randomized graph model. Thus, exact algorithms for motif analysis tend to scale very poorly with graph and pattern sizes. 
A common algorithm design pattern to speed up the computation, albeit sacrificing the quality of the solutions, is relying on approximated algorithms. In particular, sampling methods are a popular choice for the task of motif discovery.

The increasing availability of large-scale real-world datasets with group interactions calls for the development of more efficient and usable algorithms for higher-order motif discovery. 
In this work, we build on our preliminary results regarding counting all the possible patterns of higher-order interactions involving a given number of nodes~\cite{lotito2021higher}, and propose:
\begin{itemize}
    \item an efficient exact algorithm for performing higher-order motif analysis with motifs involving $3$ and $4$ nodes, including efficiently solving the hypergraph isomorphism problem for small instances and constructing vertex-induced sub-hypergraphs;
    \item an approximated method based on hyperedge sampling that overcomes the scalability issues of the exact algorithm at the expense of only a very limited decrease of accuracy;
    \item the application of the approximated method to the study of higher-order motifs of order $5$ (which are generally intractable for exact algorithms) in several networks of interest. 
\end{itemize}

This paper is organized as follows. In~\Cref{sec:related} we survey related work. In~\Cref{sec:preliminaries} we introduce basic definitions and formalize the problem of interest. In~\Cref{sec:algorithms} we outline our proposals for solving exactly the problem of higher-order motif discovery in hypergraphs. In~\Cref{sec:sampling} we propose a sampling algorithm for the same problem. In~\Cref{sec:eval} we evaluate our proposals. In~\Cref{sec:conclusion} we conclude the paper. 

\section{Related work}
\label{sec:related}
Network motifs have been extensively investigated:
\begin{itemize}
    \item in the field of network science, for their relevance in the study of the local structure and the interplay between the topology and the dynamics of complex networks;
    \item in the field of data mining, due to the complexity of the problem of enumerating connected subgraphs up to a certain size from large graphs.
\end{itemize}
In this section, we propose a brief survey of the relevant prior work in both areas and highlight our contributions to both fields.

\subsection{Network science}
Network motifs describe complex networks by their preferential patterns of interactions at the microscale. They can be interpreted as fundamental circuits that have a role in the functionality of a system~\cite{milo2002network,schwarze2020motifs}, and are therefore able to discriminate networks that represent systems from different domains or with different functionalities~\cite{milo2004superfamilies}. 
The notion of network motifs has been extended to a variety of generalized network models, to encode and quantify a pattern of interactions with a richer set of features. 
Onnela et al. generalize network motifs to weighted networks, characterizing subgraphs in terms of their intensity and coherence~\cite{onnela2005intensity}.
In temporal networks, the topological and temporal microscale is described in terms of patterns of interactions inside restricted time windows~\cite{kovanen2011temporal,paranjape2017motifs}.
Battiston et al. extend motif analysis to multilayer networks by considering over-expressed subgraphs spanning across several layers~\cite{battiston2017multilayer}.
More recently, motif analysis has been extended also to consider patterns of higher-order interactions, i.e., interactions involving an arbitrary number of nodes. 
In particular, Lee et al. characterize real-world hypergraphs at their microscale in terms of patterns of connected hyperedges of fixed size~\cite{lee2020hypergraph}. However, they do not follow the traditional approach to network motifs first proposed by Milo et al.~\cite{milo2002network}.

\subsection{Data mining} 
Being the task of motif discovery of practical utility but computationally expensive (equivalent to the subgraph isomorphism problem, which is known to be NP-complete), a vast amount of literature has been developed around this problem. 
In particular, the development of faster algorithms for motif analysis has been historically motivated mainly by the ever-growing size of biological datasets to be analyzed. 
The algorithms for motif discovery can be clustered in two big groups: \textit{exact counting}  and \textit{sampling methods}. 
Each type of algorithm comes with advantages and drawbacks: exact algorithms allow extracting the correct number of occurrences of each motif; however, they are usually slow and memory-intensive. 
Sampling methods are generally faster, but they need to be designed to avoid biases and only offer an approximation of the count of each motif. 
Historically, the interest of the community has moved from developing exact counting algorithms to developing approximated algorithms and heuristics able to tackle larger networks and extract larger motifs in an acceptable time, in spite of losing accuracy. 
The very first motif discovery algorithm was the exhaustive enumeration method proposed by Milo et al.~\cite{milo2002network}. 
The enumeration of all the possible patterns of subgraphs of a given number of nodes is computationally expensive, and only really small subgraphs of size $3$ and $4$ could be analyzed. 
After that, a plethora of improvements have been proposed. Among those, we highlight the ESU and RAND-ESU algorithms~\cite{Wernicke2006Efficient}. 
ESU is an efficient algorithm that enumerates the set of all induced subgraphs of a given size from a large graph exploiting a tree-like data structure.
RAND-ESU is an unbiased sampling method built on top of ESU that samples the branches of the tree to visit. For a more in-depth overview of these algorithms, the interested reader can refer to~\cite{Sabyasachi2020review}.

While the amount of research for classic network motifs is huge, the data mining literature for generalized network motifs is far from being as complete, with some exceptions for temporal networks~\cite{liu2019sampling,Jazayeri2020motif}. 
In particular, the computational aspects associated with motif discovery in hypergraphs are largely overlooked. 
In their seminal paper~\cite{horvat2007frequent}, Horv{\'a}th et al. propose an incremental polynomial time algorithm for the related problem of mining frequent sub-hypergraphs in
hypergraph databases. 
In their setting, the frequency of a sub-hypergraph corresponds to the number of hypergraphs in the database containing a query sub-hypergraph. 
In the case of mining motifs, in which only one input hypergraph is considered, this algorithm can only find if a certain pattern of sub-hypergraph is present or not in the input hypergraph. 
Therefore, the algorithm proposed by Horv{\'a}th et al. is not suited for mining motifs in a hypergraph, since we want to compute the exact number of occurrences of each possible pattern of sub-hypergraph in a given input hypergraph. 
Preti et al. tackle the problem of mining frequent patterns in simplicial complexes~\cite{preti2022fresco}. Albeit both hypergraphs and simplicial complexes can represent higher-order interactions, the two mathematical objects present an important difference: simplicial complexes respect the downward closure property (i.e., in the case of hypergraphs, if a hyperedge $e$ exists in a hypergraph $\mathcal {H}$ then also all the proper subsets of $e$ exists in $\mathcal {H}$).
Lee et al. propose a set of algorithms, including an exact, a parallel and an approximated algorithm for mining motifs in hypergraphs~\cite{lee2020hypergraph,lee2021thyme+}. However, they focus only on the problem of extracting patterns of overlaps between hyperedges of fixed size.

\subsection{Contributions}

In our previous work~\cite{lotito2021higher}, we adopt a traditional approach and extend to hypergraphs the original network motif definition proposed by Milo et al.~\cite{milo2002network}, investigating all possible patterns of pairwise and group interactions among a given number of nodes. In~\cite{lotito2021higher}, we extract fingerprints of real-world hypergraphs at their microscale and identify key motifs associated with families of hypergraphs belonging to different domains. Here, we build on our first work and further develop our algorithm for the efficient computation of frequencies of small patterns of sub-hypergraphs. While our proposal can outperform the baseline in every dataset, exact methods suffer from high computational costs and do not scale to large hypergraphs nor to patterns beyond four nodes. Hence, this paper introduces an approximated method for higher-order motif extraction based on hyperedge sampling as well. We show that our sampling method dramatically speeds up computations at the cost of only minimal errors in the estimation of motif frequency and allows us to analyse larger higher-order motifs beyond the computational limits of the exact approach.

\section{Preliminaries and problem statement}\label{sec:preliminaries}
Here we recall some basic definitions and give a formal description of the problem of our interest.

\begin{definition}[Hypergraph]
A hypergraph is a pair $\mathcal{H} = (V, E)$ where $V$ is the set of the vertices and $E$ is the set of the \emph{hyperedges}. A hyperedge $e$ is a subset of $V$ linking the vertices contained in it. 
\end{definition}

\begin{figure*}[t]
    \centering
    \includegraphics[scale=0.5]{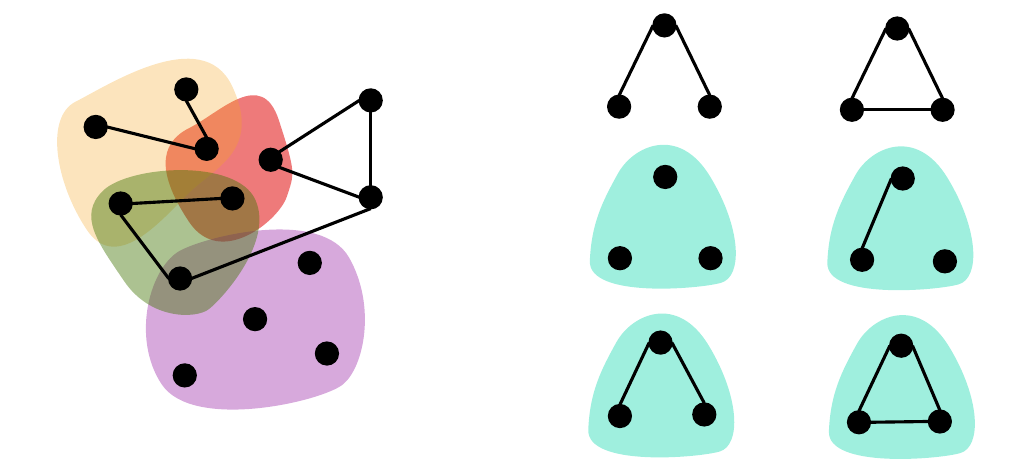}
    \caption{On the left, an example of a small higher-order network represented by a hypergraph. On the right, we show all the possible patterns of higher-order interactions involving three connected nodes.}
    \label{fig:enum_3nodes}
\end{figure*}

In~\Cref{fig:enum_3nodes} we show an example of a hypergraph. We recall that a hypergraph in which every hyperedge links two vertices corresponds to the standard definition of a graph.

\begin{definition}[Vertex-induced sub-hypergraph]
The sub-hypergraph $\mathcal{H}[V']$ induced by the subset $V' \subseteq V$ is the pair $(V', E')$, where $E' = \{e \in E : e \subseteq V'\}$.
\end{definition}

Now we can define more formally the notions related to the \emph{isomorphism} problem, which is the fundamental theoretical tool underlying network motif discovery.

\begin{definition}[Hypergraph isomorphism]
Two hypergraphs $\mathcal{H}=(V,E)$ and $\mathcal{H}'=(V',E')$ are isomorphic if they are identical modulo relabeling of the vertices. More formally, if there exists a bijection $f: V \rightarrow V'$ such that  $e = \{u_1, ..., u_n\} \in E$ if and only if $e' = \{f(u_1), ..., f(u_n)\} \in E'$.
\end{definition}

\begin{definition}[Occurrence]
Given a hypergraph $\mathcal{H}=(V,E)$ and a smaller query hypergraph $\mathcal{Q}=(V', E')$, the occurrences of $\mathcal{Q}$ in $\mathcal{H}$ are all the sub-hypergraphs of $\mathcal{H}$ isomorphic to $\mathcal{Q}$. We often refer to the number of occurrences of $\mathcal{Q}$ in $\mathcal{H}$ as the \emph{frequency} of $\mathcal{Q}$ in $\mathcal{H}$.
\end{definition}

Finally, we can give a formal description of the problem of our interest.

\begin{definition}[Higher-order motifs]
Higher-order motifs are patterns of sm\-all connected sub-hypergraphs that occur in an observed hypergraph $\mathcal{H}$ at a frequency that is significantly higher with respect to a null model. We refer to the number of nodes involved in the pattern as the \emph{order} of a higher-order motif.
\end{definition}

As previously mentioned, to perform a higher-order motif analysis of a system, one needs to (i) count the frequency of each query higher-order motif in a hypergraph, (ii) compare the frequency of each query higher-order motif with that observed in a null model, and (iii) evaluate the over- or under-expression of query each higher-order motif. In this work we are mostly interested in the first step, i.e., motif discovery, therefore we will use higher-order motifs to refer to all the possible patterns of sub-hypergraphs involving a certain number of nodes. In~\Cref{fig:enum_3nodes} we enumerate all the higher-order motifs of order $3$. 

\paragraph*{Problem (Mining higher-order motifs).}
%\begin{problem}[Mining higher-order motifs]
Given a hypergraph $\mathcal{H}$ and an integer $k>2$, compute the frequency of every higher-order motif of order $k$.
\label{problem:mining}
%\end{problem}

\section{Mining higher-order motifs}
\label{sec:algorithms}
The enumeration of all the patterns of connected sub-hypergraphs of a given size is obviously the most expensive sub-task in motif analysis. The weight of this step is even more impactful considering that it must be repeated in randomized networks as well. To solve this problem exactly, in this section, we propose a baseline algorithm based on projecting the hypergraph onto a graph and employing state-of-the-art motif analysis algorithms on it. Additionally, we present a more efficient method that directly leverages higher-order structures to construct sub-hypergraphs of a specified size.

\subsection{Baseline algorithm}
While traditional algorithms are not able to identify patterns of polyadic interactions, they can be used as a routine for more sophisticated algorithms. In our baseline, we consider the \emph{projected graph} of a hypergraph. 

\begin{definition}[Projection of a hypergraph]
The projection of a hypergraph $\mathcal{H}=(V,E)$ is a graph $G = (V, E')$, defined on the same vertices of $\mathcal{H}$ and such that an edge between two vertices $a, b \in V$ exists if and only if $a$ and $b$ participate together in at least a hyperedge $e \in E$. In other words, every hyperedge $e \in E$ is replaced in $G$ with a clique. 
\end{definition}
  
\begin{figure}
    \centering
    \includegraphics[width=\linewidth]{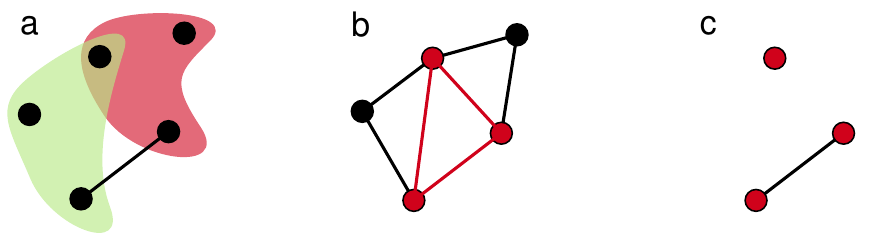}
    \caption{a) Example of a hypergraph $H$ in which the baseline fails. b) We highlight in red a connected subgraph $s$ of size $k=3$, one of the many possible outputs of a standard motif discovery algorithm applied on the projection of the previous hypergraph. c) The sub-hypergraph induced by the vertices of $s$ and the hyperedges of $H$ is not connected.}
    \label{fig:baseline_fig}
\end{figure}

By running a classic algorithm (e.g., ESU~\cite{Wernicke2006Efficient}), we can efficiently enumerate connected subgraphs of size $k$ in the projected graph. However, these subgraphs are only \emph{candidate higher-order motifs} for two potential reasons: (i) they do not include higher-order interactions; (ii) even if a subgraph $s$ of size $k$ is connected in the projected graph, the sub-hypergraph induced by the vertices in $s$ and the hyperedges $E$ may be not connected. We highlight these pitfalls in~\Cref{fig:baseline_fig}.

In order to account for these issues, we construct the sub-hypergraph induced by the $k$ nodes of the candidate motif (see~\Cref{subsec:details}) and check if this sub-hypergraph is connected. If it is, then one can simply update the frequency hash map, otherwise, the output is discarded. A more formal explanation of this method is reported in~\Cref{alg:motif_count}. All in all, the baseline inherits the complexity of ESU~\cite{Wernicke2006Efficient}), plus the preprocessing cost of computing the clique projection of the hypergraph.

\begin{algorithm} 
\caption{Baseline: Counting higher-order motifs}
\label{alg:motif_count}
\begin{algorithmic}[1]
\renewcommand{\algorithmicrequire}{\textbf{Input:}}
\Require a hypergraph $\mathcal{H} = (V,E)$ and an integer $k \in \{3, 4\}$.
\renewcommand{\algorithmicrequire}{\textbf{Output:}}
\Require distribution of the frequency of the motifs of order $k$.

\State Let $G = (V,E')$ be the projection of $\mathcal{H}$
\State Let $\mathcal{M}$ be the motifs frequency dictionary
\State $\mathcal{S} \gets \Call{ESU}{G, k}$
\ForEach{$\textit{subgraph} = (V^*,E^*) \in \mathcal{S}$}
\State \textit{candidate\_motif} $\gets$ sub-hypergraph (of $\mathcal{H}$) induced by $V^*$
\If{\textit{candidate\_motif} is connected}
\State Let $\mathcal{C}_m$ be the isomorphism class of \textit{candidate\_motif}
\State $\mathcal{M}[\mathcal{C}_m] \textrm{~+=~} 1$
\EndIf

\EndFor
\end{algorithmic}
\end{algorithm}

\subsection{Efficient algorithms}
The most expensive step in the previous algorithm is obviously the ESU subroutine. Moreover, the performance is widely impacted by the fact that hypergraph projections can be very dense and that lots of subgraphs are discarded for not satisfying the requirement of the induced sub-hypergraphs of being connected. To solve these problems, we work directly on hypergraphs, designing an efficient algorithm that exploits containment properties of higher-order structures in real-world systems. We optimize separately the two cases of $3$- and $4$-node motifs.

\paragraph*{$3$-node motifs}
As shown in~\Cref{fig:enum_3nodes}, two of the motifs involving three nodes are composed only by pairwise relations, while the others involve one hyperedge of order $3$. To discover the latter, it is enough to iterate over all the hyperedges of order $3$ and then recover the nested pairwise links to build the motif (``fill in'' the hyperedges, see~\Cref{subsec:details}); the sub-hypergraph is trivially connected since its nodes are part of the same hyperedge. Then, we can ignore all the higher-order interactions and focus only on the pairwise links, since we are interested in counting the frequency of the first two motifs of~\Cref{fig:enum_3nodes}. In this case, we can rely on ESU. This time, however, it will need to handle a lot fewer edges. Every time ESU returns an output, the triplet of nodes could have been counted already in the previous step (i.e., overlap between a pairwise motif and a hyperlink of order $3$): in this case, the triplet is discarded. The first step has a complexity linear in the number of the hyperedges of size $3$, while the second step inherits the complexity of the ESU algorithm.
A formal description of the algorithm for higher-order motifs of order $3$ is reported in~\Cref{alg:motif_count_3}.

\begin{algorithm} 
\caption{Efficient algorithm: Counting higher-order motifs of order $3$}
\label{alg:motif_count_3}

\begin{algorithmic}[1]
\renewcommand{\algorithmicrequire}{\textbf{Input:}}
\Require a hypergraph $\mathcal{H} = (V,E)$.
\renewcommand{\algorithmicrequire}{\textbf{Output:}}
\Require distribution of the frequency of the motifs of order $3$.

\State Let $\mathcal{M}$ be the motifs frequency hash map
\ForEach{hyperedge $e$ of order $3$ in $E$}
\State $V^* \gets$ vertices of $e$
\State \textit{motif} $\gets$ sub-hypergraph induced by $V^*$
\State Let $\mathcal{C}_m$ be the isomorphism class of \textit{motif}
\State $\mathcal{M}[\mathcal{C}_m] += 1$
\EndFor
\State $G \gets$ Discard all hyperedges of order $3$ from $\mathcal{H}$
\State $\mathcal{S} \gets \Call{ESU}{G, 3}$
\ForEach{$\textit{subgraph}=(V^*,E^*) \in \mathcal{S}$}
\If{$V^*$ not already visited}
\State Let $\mathcal{C}_m$ be the isomorphism class of $\textit{subgraph}$
\State $\mathcal{M}[\mathcal{C}_m] += 1$
\EndIf
\EndFor
\end{algorithmic}
\end{algorithm}

\paragraph*{$4$-node motifs}
The algorithm for motifs of order $4$ is similar, albeit there are some more details to take into account. One can still iterate on all the hyperedges of order $4$, count the motifs by considering also the rich nested structures (one can observe that this time also hyperedges of order $3$ can be nested), and discard all the $4$-hyperedges. However, as a second step, one needs also to iterate over all the $3$-hyperedges and consider all the possible neighbours; in fact, $3$-hyperedges define only a sub-hypergraph with $3$ nodes, while we are requesting $4$ nodes. Neighbours can be listed by considering all the edges that add only $1$ new node since $3$ nodes are already fixed. The last step is to consider only pairwise interactions, and we rely again on ESU. Here, again, the first step has a complexity linear in the number of hyperedges of size $4$, the second step has a complexity quadratic in the total number of hyperedges (linear in the number of hyperedges of size $3$ and then linear for each hyperedge to explore its neighbourhood), and the final step inherits the complexity of the ESU algorithm. A formal description of the algorithm for higher-order motifs of order $4$ is reported in~\Cref{alg:motif_count_4}.

\begin{algorithm} 
\caption{Efficient algorithm: Counting higher-order motifs of order $4$}
\label{alg:motif_count_4}

\begin{algorithmic}[1]
\renewcommand{\algorithmicrequire}{\textbf{Input:}}
\Require a hypergraph $\mathcal{H} = (V,E)$.
\renewcommand{\algorithmicrequire}{\textbf{Output:}}
\Require distribution of the frequency of the motifs of order $4$.

\State Let $\mathcal{M}$ be the motifs frequency hash map
\ForEach{hyperedge $e$ of order $4$ in $E$}
\State \textit{motif} $\gets$ sub-hypergraph induced by $e$
\State Let $\mathcal{C}_m$ be the isomorphism class of \textit{motif}
\State $\mathcal{M}[\mathcal{C}_m] \textrm{~+=~} 1$
\State Set vertices of \textit{motif} as visited
\EndFor
\State $\mathcal{H} \gets$ Discard all hyperedges of order $4$ from $\mathcal{H}$
\ForEach{hyperedge $e$ of order $3$ in $E$}
\State Let $\mathcal{E}$ be the set of hyperedges adjacent to $e$
\ForEach{hyperedge $e_i$ in $\mathcal{E}$}
\If{$\lvert e \cup e_i \rvert = 4$ and $e \cup e_i$ not already visited}
\State \textit{motif} $\gets$ sub-hypergraph induced by $e \cup e_i$
\State Let $\mathcal{C}_m$ be the isomorphism class of \textit{motif}
\State $\mathcal{M}[\mathcal{C}_m] \textrm{~+=~} 1$
\State Set vertices of \textit{motif} as visited
\EndIf
\EndFor
\EndFor
\State $\mathcal{H} \gets$ Discard all hyperedges of order $3$ from $\mathcal{H}$

\State $\mathcal{S} \gets \Call{ESU}{\mathcal{H}, 4}$
\ForEach{$\textit{subgraph}=(V^*,E^*) \in \mathcal{S}$}
\If{$V^*$ not already visited}
\State Let $\mathcal{C}_m$ be the isomorphism class of $\textit{subgraph}$
\State $\mathcal{M}[\mathcal{C}_m] \textrm{~+=~} 1$
\State Set $V^*$ as visited
\EndIf
\EndFor
\end{algorithmic}
\end{algorithm}

\subsection{Algorithm details}
\label{subsec:details}
\begin{figure}
    \centering
    \includegraphics[width=\linewidth]{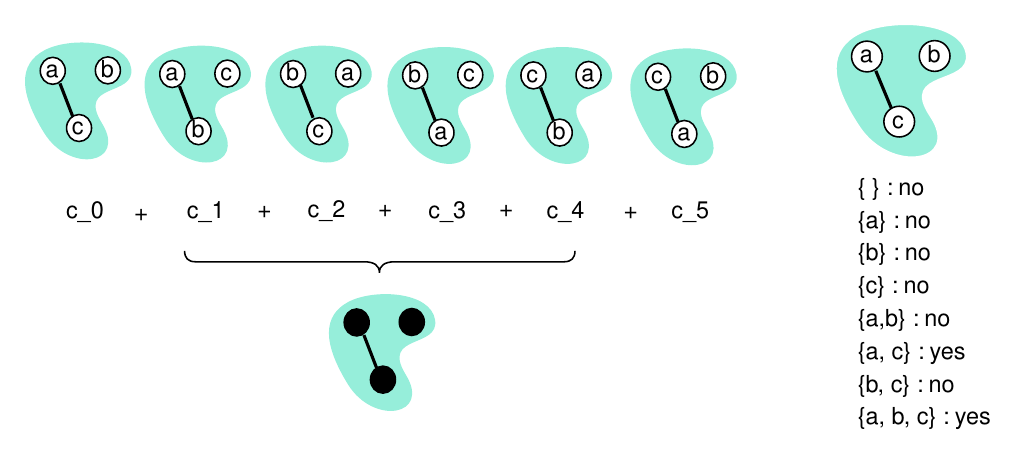}
    \caption{a) On the left, we show how to efficiently solve the problem of hypergraph isomorphism for small hypergraphs. We generate and hash every possible pattern of higher-order interactions involving $k$ nodes with all the corresponding relabelings. Every observed sub-hypergraph will be equivalent to one and only one of the entries of the hash map. The final count of each motif is the sum of all the entries of the hash map that belong to the same isomorphism class. b) On the right, we show how to construct vertex-induced sub-hypergraphs efficiently. As a preprocessing step, we hash every hyperedge in a hypergraph, allowing us to check for their existence in constant time. For a query set of $3$ or $4$ vertices, we generate all the possible $2^3$ or $2^4$ subsets of the query set and check in constant time if each subset is an existing hyperedge. Every time a subset is found to exist, we add it to the sub-hypergraph induced by the query set.}
    \label{fig:algorithm_details}
\end{figure}

Counting higher-order motifs can be interpreted as the enumeration of all the possible connected sub-hypergraphs of size $k$, assigning each of them to an isomorphism class. An efficient way to assign an isomorphism class to a connected sub-hypergraph of size $k$ (for small values of $k$) is relying on a hash map. One can generate and hash every possible pattern of higher-order interactions involving $k$ nodes, with all the possible relabelings. Relabelings are important because the same sub-hypergraphs can be stored with different labels on the vertices. For example, we have $6$ different patterns of higher-order interactions with $3$ nodes, each with $3!$ possible relabeling; eventually, the hash map will contain $6 \cdot 3! = 36$ entries. One can use the hash map as a counter since each observed sub-hypergraph is a key. After having enumerated all the sub-hypergraphs, the final count of each motif is simply the sum of all the entries of the hash map that belong to the same isomorphism class. We show a summary of this process in~\Cref{fig:algorithm_details}. Considering the sizes of the sub-hypergraphs involved, we can assume that this process incurs a constant time cost.

Another important routine in our algorithms is the construction of vertex-induced sub-hypergraphs. Given a set of vertices $V'$, we are interested in querying the set of all the hyperedges to extract those who have all their endpoints in $V'$. This is what we referred to as ``filling in'' a set of vertices in the previous sections. For our specific case, this problem is efficiently solvable relying again on hash maps as follows. We can hash every hyperedge of a hypergraph: this ensures that we are able to check the existence of a hyperedge in constant time. Since we are only interested in solving this problem for a query set of vertices of size $3$ or $4$, we can easily generate all the possible $2^3$ or $2^4$ subsets of vertices (we can also ignore the empty set and the singletons) and check in constant time if each subset is an existing hyperedge. We show a summary of this process in~\Cref{fig:algorithm_details}. All in all, given that we are interested in very small sets of vertices, we can construct vertex-induced sub-hypergraphs in constant time.

\section{Sampling methods for counting higher-order motifs}
\label{sec:sampling}

Scalability is a persistent issue for exact motif discovery algorithms. Motif analysis has a number of real-world applications that require handling vast datasets. However, exact algorithms for motif discovery quickly become intractable for realistic inputs and motif sizes. To address this complexity, we propose an approximated method based on hyperedge sampling.

\Cref{alg:sampling_motif_count_4} samples with replacement $S$ times a hyperedge $e$ and enumerates all the connected sub-hypergraphs with a given number of nodes and containing $e$. The number of samples $S$ controls the quality of the approximated results. However, directly sampling hyperedges from the hypergraphs leads to unreliable results. The distribution of hyperedge sizes is non-uniform, causing the algorithm to often sample hyperedges of size $2$ while seldom sampling those of size $4$. This skews the estimation of specific sub-hypergraph patterns. To mitigate this, we employ stratified sampling, segmenting the sampling process to guarantee a balanced consideration of hyperedges across different sizes. Let $S_k$ be the number of samples assigned to hyperedges of size $k$, such that $S = \sum_k S_k$. We estimate appropriate values for $S_k$ for every $k$ empirically, exhaustively searching among different combinations of values and selecting those that maximize a defined quality function (see~\Cref{appendix}).

The sampling algorithm proceeds in a way similar to the exact method (therefore we avoid explicitly repeating some details in the pseudocode). If we target the discovery of motifs of size $k$, then for all the sampled hyperedges of size $k$ we have all the necessary vertices to build a target sub-hypergraph. No exploration of the neighbourhood is further required to add new nodes to the pattern. The complexity of this step is linear in the number of samples $S_k$. Then, for all the sampled hyperedges of size less than $k$, some exploration of possibly different levels of the neighbourhoods is required. The complexity of each of these steps is linear in the number of samples $S_k$ multiplied, for each level of exploration, by a factor linear in the number of hyperedges. Moreover, for each pattern, the previously mentioned process of ``filling in'' the hyperedges is repeated to build vertex-induced sub-hypergraphs and count the right instances of the motifs. Again, these routines take constant time.

In order to estimate the exact count for each motif, the algorithm multiplies the observed count by a correction factor given by the probability of sampling a certain motif, as reported in the pseudocode. To simplify the computation of the correction factor, the algorithm discards all sub-hypergraphs encountered during the exploration of the neighbourhood of a hyperedge $e$ that contains at least one hyperedge with a higher cardinality than $e$. In other words, a pattern of sub-hypergraph is only considered when the hyperedge of maximal cardinality is sampled. Given this approach, it is straightforward to prove that the estimator is unbiased.

\begin{algorithm} 
\caption{Sampling algorithm: Counting higher-order motifs of order $4$}
\label{alg:sampling_motif_count_4}

\begin{algorithmic}[1]
\renewcommand{\algorithmicrequire}{\textbf{Input:}}
\Require a hypergraph $\mathcal{H} = (V,E)$, number of hyperedge to sample $S$.
\renewcommand{\algorithmicrequire}{\textbf{Output:}}
\Require approximated distribution of the frequency of the motifs of order $4$.
\State Let $\mathcal{M}$ be the motifs frequency hash map
\State Let $E_k$ represent the set of hyperedges of $\mathcal{H}$ of size $k$ 
\State Let $S_k$ be the number of samples assigned to hyperedges of size $k$ (ensure that the sum of the $S_k$ for every possible $k$ is equal to $S$)
\State $\textit{sampled\_edges} \gets $ sample $S_2$ hyperedges from $E_2$
\State $\textit{sampled\_edges} \gets \textit{sampled\_edges } \cup$ sample $S_3$ hyperedges from $E_3$
\State $\textit{sampled\_edges} \gets \textit{sampled\_edges } \cup$ sample $S_4$ hyperedges from $E_4$
\ForEach{hyperedge $e$ in $\textit{sampled\_edges}$}
\State Let $\mathcal{S}$ be the set of connected sub-hypergraphs of $\mathcal{H}$ containing $e$
\ForEach{\textit{sub-hypergraph} in $\mathcal{S}$}
\If{\textit{sub-hypergraph} contains a hyperedge of cardinality bigger than $e$}
\State continue
\EndIf
\State Let $\mathcal{C}_m$ be the isomorphism class of \textit{sub-hypergraph}
\State $\mathcal{M}[\mathcal{C}_m] \textrm{~+=~} 1$
\EndFor
\EndFor

\ForEach{motif $m$ with count $c$ in $\mathcal{M}$}
\State Let \textit{maxcard} be the maximum size of the hyperedges in $m$
\State Let \textit{countmax} be the number of the hyperedges in $m$ of size \textit{maxcard}
\State $c \gets \frac{c \cdot \lvert E_{\textit{maxcard}} \rvert}{S_{\textit{maxcard}} \cdot \textit{countmax}}$
\EndFor
\end{algorithmic}
\end{algorithm}

\section{Experimental evaluation}
\label{sec:eval}
To assess the improvement in the performance of the algorithms for higher-order motif discovery when (i) exploiting higher-order structures instead of applying classic methods on the hypergraph projection and (ii) approximating motif frequency, we collected a variety of hypergraph datasets representing real-world systems with group interactions. Besides the evaluation of the performance, we also study the accuracy of the sampling algorithm and exploit sampling methods to study higher-order motifs of order $5$. All the experiments have been carried out on a machine with an 8-core (2.2GHz) Intel Xeon CPU and 94GB of RAM, running Ubuntu 20.04.4 LTS. The algorithms presented in this paper are implemented in Python3. The code is publicly available~\cite{code2022sampling}. Moreover, all the algorithms presented in this work are included in the Python library Hypergraphx for higher-order network analysis~\cite{lotito2023hypergraphx}. 

\subsection{Datasets}
We collected a variety of real-world datasets from different domains, describing face-to-face interactions, co-authorship relations and e-mail communications. Co-authorship data (\texttt{dblp}, \texttt{history} and \texttt{geology})~\cite{benson2018simplicial} is naturally encoded as a set of nodes (the authors) involved in higher-order interactions (the scientific papers). Also E-mail data (\texttt{EU})~\cite{benson2018simplicial} is naturally encoded as a set of higher-order interactions since e-mails can have multiple recipients at the same time. However, higher-order interactions need to be inferred from pairwise relations in data about face-to-face interactions (\texttt{primary school} and \texttt{high school})~\cite{benson2018simplicial}. In this case, cliques of size $k$ are promoted to group interactions of order $k$, if the corresponding dyadic encounters happened at the same time. The summary statistics of the datasets are reported in Table~\ref{tab:datasets}. The datasets, as well as the preprocessing scripts, are publicly available~\cite{code2022sampling}.

\begin{table*}
\caption{Summary statistics of the datasets considered for our experiments. Each higher-order network is described by the domain, the number of nodes, and the total number of hyperedges of size $2$, $3$, $4$ and $5$.}
\centering
\begin{tabular}{||l l r r r r r||}
\toprule
Dataset & Domain & N & $E_2$ & $E_3$ & $E_4$ & $E_5$\\
\midrule
\texttt{hs} & \texttt{proximity} & 327 & 5498 & 2091 & 222 & 7\\
\texttt{ps} & \texttt{proximity} & 242 & 7748 & 4600 & 347 & 9\\
\texttt{EU} & \texttt{e-mail} & 998 & 12753 & 4938 & 2294 & 1359\\
\texttt{dblp} & \texttt{co-auth} & 1924991 & 693363 & 667291 & 419434 & 205970\\
\texttt{history} & \texttt{co-auth} & 1014734 & 160885 & 47423 & 19120 & 8775\\
\texttt{geology} & \texttt{co-auth} & 1256385 & 275736 & 227950 & 159509 & 99140\\

\bottomrule
\end{tabular}
\label{tab:datasets}
\end{table*}

\subsection{Performance evaluation}

\begin{table*}
\caption{Comparison of the running time (s) of the exact algorithms with motifs of order $3$ and $4$.}
\centering
 \begin{tabular}{||p{30mm} p{15mm} p{15mm} p{15mm} p{15mm}||}
\toprule
 Dataset & Base-3 & Eff-3 & Base-4 & Eff-4\\ [0.5ex]
 \midrule
 \texttt{hs} & 7 & \textbf{5} & 362 & \textbf{230} \\
 \texttt{ps} & 25 & \textbf{18} & 1920 & \textbf{1339}\\
 \texttt{EU} & 44 & \textbf{29} & 5286 & \textbf{2757}\\
   \texttt{dblp} & 1185 & \textbf{134} & $> 24$h & \textbf{2885}\\
    \texttt{history} & 42 & \textbf{19} & 4591 & \textbf{526}\\
    \texttt{geology} & 207 & \textbf{36} & 32810 & \textbf{475}\\
  \bottomrule
\end{tabular}
\label{table:experiments}
\end{table*}

In~\Cref{table:experiments} we compare our exact algorithm for higher-order motif discovery against the baseline algorithm in terms of their execution running time. We show that exploiting directly higher-order structures speeds up the computations. The efficient algorithm outperforms the baseline in every dataset. Moreover, it is worth mentioning that the analysis with motifs of order $4$ of \texttt{dblp} with the baseline algorithm was not feasible in a reasonable amount of time. The larger gains are observed in co-authorship data. Co-authorship systems are proven to display a nested structure of hyperedges made up of a small number of hyperedges of large average size~\cite{lotito2021higher}. In fact, these kinds of systems are the ideal scenario for our algorithm. We can notice that the gains are not as noticeable in social datasets, which tend to be governed by dense patterns of lower-order interactions~\cite{lotito2021higher}.

In~\Cref{tab:results} we show the execution running time in seconds of the sampling algorithm on the different datasets with multiple values of $S$, i.e., the parameter that controls the number of samples. The different size scales of the co-authorship and social datasets require different sample sizes to achieve results of comparable quality. Since the analysis of the motifs of order $3$ was already easily doable, we consider only the task of motif discovery of order $4$. We show that hyperedge sampling dramatically improves performance. As expected, the parameter $S$ heavily affects the running time. As always, there is a trade-off between the accuracy of the results (higher values of $S$ lead to more accurate estimates) and the execution running time.

\begin{table*}
    \centering
    \caption{Hyperedge sampling dramatically improves the performance with respect to the exact algorithm. The execution running time of the approximated algorithm heavily depends on the choice of the sample size $S$. The correlation coefficient $\rho$ between the estimated and the exact motif profiles, the maximum absolute error MaxAE and the mean average error MAE improve with increases in the number of samples $S$. Due to their different size scale, co-authorship and social datasets require different sample sizes to achieve comparable results. We obtain reasonable results even with a very limited number of samples.}
    \begin{tabular}{ ||p{34px} c c c c c c|| } 
    \toprule
    Dataset & \makecell{Exact\\exec. time (s)} & $S$ & \makecell{Approx.\\exec. time (s)} & $\rho$ & MaxAE & MAE \\ [0.5ex]
    \midrule
    \multirow{4}{*}{\texttt{hs}} & \multirow{4}{*}{230} & 100 & 3 & .914 & .151 & .015 \\ 
    & & 250 & 8 & .953 & .122 & .011 \\ 
    & & 500 & 16 & .978 & .096 & .007 \\
    & & 1K & 29 & .987 & .065 & .006 \\
    \hline
        \multirow{4}{*}{\texttt{ps}} & \multirow{4}{*}{1339} & 100 & 11 & .918 & .151 & .017\\ 
    & & 250 & 30 & .950 & .135 & .012\\ 
    & & 500 & 63 & .977 & .093 & .008\\
    & & 1K & 118 & .986 & .071 & .006\\
    \hline
        \multirow{4}{*}{\texttt{EU}} & \multirow{4}{*}{2757} & 100 & 15 & .804 & .203 & .028\\     & & 250 & 34 & .887 & .159 & .020\\ 
    & & 500 & 73 & .923 & .134 & .016\\
    & & 1K & 144 & .963 & .098 & .010\\
    \hline
        \multirow{4}{*}{\texttt{dblp}} & \multirow{4}{*}{2885} & 1K & 32 & .495 & .088 & .055\\ 
    & & 2.5K & 41 & .555 & .088 & .047\\ 
    & & 5K & 56 & .610 & .089 & .038\\
    & & 10K & 85 & .696 & .087 & .027\\
    \hline
        \multirow{4}{*}{\texttt{history}} & \multirow{4}{*}{526} & 1K & 5 & .679 & .176 & .033\\ 
    & & 2.5K & 7 & .804 & .169 & .022\\ 
    & & 5K & 11 & .867 & .170 & .016\\
    & & 10K & 22 & .913 & .124 & .012\\
    \hline
        \multirow{4}{*}{\texttt{geology}} & \multirow{4}{*}{475} & 1K & 12 & .590 & .107 & .047\\ 
    & & 2.5K & 15 & .661 & .107 & .039 \\ 
    & & 5K & 19 & .719 & .106 & .032 \\
    & & 10K & 30 & .784 & .103 & .024 \\
    \bottomrule
    
    \end{tabular}

    \label{tab:results}
\end{table*}

\subsection{Accuracy of sampling method} Besides evaluating the running time of the sampling method, it is also important to assess the output quality of the estimates compared to exact higher-order motif profiles. We compute motif profiles~\cite{lotito2021higher} comparing the observed frequencies of the motifs with those on a null model~\cite{chodrow2020configuration} to assess their statistical significance (we sample $N=10$ times from the configuration model).

We evaluate the quality of the estimated motif profiles in terms of:

\begin{itemize}
    \item the \emph{Pearson's correlation coefficient $\rho$} between the estimated and the exact higher-order motif profiles. The coefficient assigns values close to $1$ to profiles in strong agreement and values close to $-1$ to profiles in strong disagreement.
    \item the \emph{Maximum Absolute Error (MaxAE)} in estimating higher-order motif profiles.
    \item the \emph{Mean Absolute Error (MAE)} in estimating higher-order motif profiles.
\end{itemize}

 In~\Cref{tab:results}, we show that we obtain good results even with a small number of samples with respect to the total size of the hypergraphs. The measures of output quality improve with increases in $S$. A good trade-off between the output quality, $S$ and the execution running time will be critical for real-world applications. The measures of output quality are averaged across $10$ repetitions for every value of $S$. 

\begin{figure}
    \centering
    \includegraphics[scale=0.5]{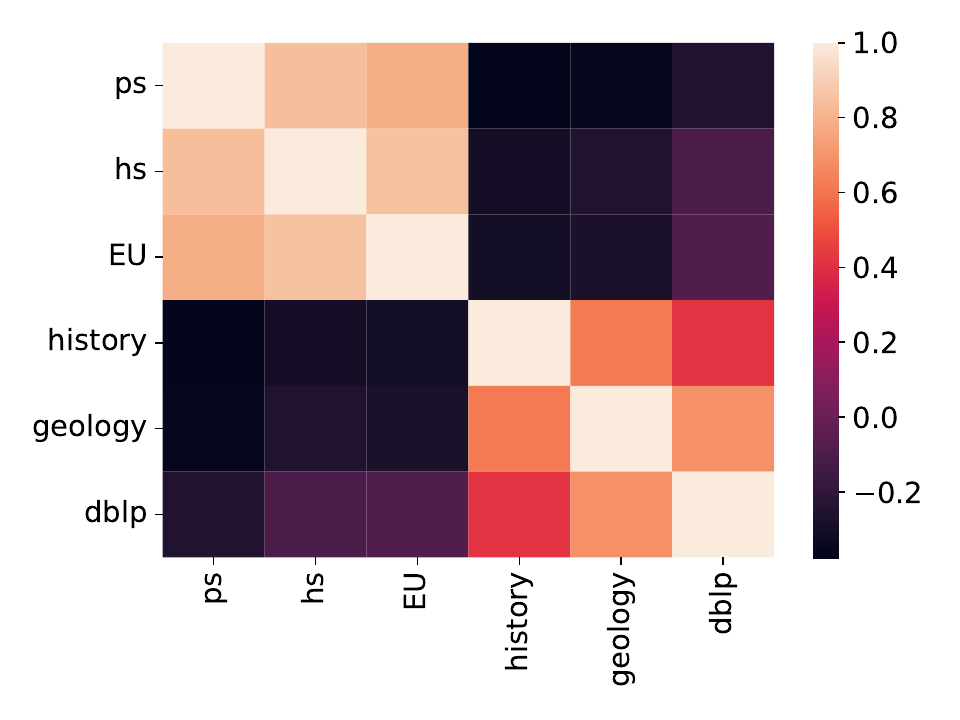}
    \caption{The correlation matrix of the significance profiles built with sampling methods ($S=1000$ for co-authorship data and $S=100$ for social data) highlights the emergence of two clusters that separate social and co-authorship data.}
    \label{fig:cluster-rand}
\end{figure}

A second evaluation metric is the correlation matrix of the motif significance profiles~\cite{milo2002network,lotito2021higher} of the different real-world hypergraphs. In~\Cref{fig:cluster-rand}, the correlation matrix shows the emergence of two ``superfamilies'' of real-world hypergraphs, in a way similar to~\cite{lotito2021higher}. Clustering tends to separate social and co-authorship data. This further proves that sampling methods are still able to capture and highlight patterns of higher-order interactions that are probably linked to the functionalities of the networks.  

\subsection{Applications: Mining larger higher-order motifs}
\begin{figure}
    \centering
    \includegraphics[width=\linewidth]{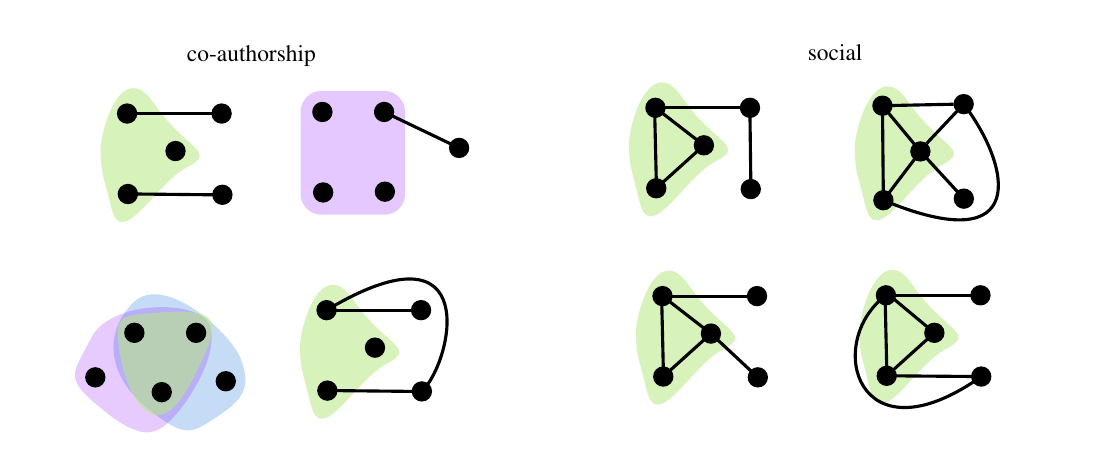}
    \caption{Over-expressed patterns of higher-order interactions highlight structural principles of the different domains.}
    \label{fig:overexpressed}
\end{figure}

Approximated methods not only speed up motif analysis for large datasets but also allow for the study of larger patterns of interactions. Exact counting algorithms are suited only for the extraction of motifs of order $3$ and $4$. Here we employ our proposed sampling algorithm and characterize two real-world hypergraphs, namely \texttt{history} and \texttt{high school}, in terms of their higher-order motifs of order $5$. To evaluate the statistical significance of the results, we compare the results with those on a configuration model. We use the same statistical evaluation methods proposed in~\cite{lotito2021higher}, i.e., we consider the relative abundance of each motif with respect to a configuration model~\cite{chodrow2020configuration}. In~\Cref{fig:overexpressed} we show the most over-expressed higher-order motifs of order $5$ in both the real-world hypergraphs. We can notice how also at this scale we still observe characteristic patterns of co-authorship data (low number of interactions of large average size) and face-to-face data (high number of interactions of small average size). This is in line with the results on the nested structure of real-world hyperedges proposed in~\cite{lotito2021higher}.

\section{Conclusion}
\label{sec:conclusion}
In this paper, we have proposed a first algorithmic framework for the problem of higher-order motif discovery in hypergraphs as defined in~\cite{lotito2021higher}. We have developed an efficient exact algorithm for the analysis of the higher-order motif of order $3$ and $4$. Such an exact algorithm exploits hyperedges and their nested structure to efficiently enumerate sub-hypergraphs of a given size. We have proved that considering directly the hypergraph structure of data outperforms traditional computational frameworks for network motifs that work on projected data. We have developed an approximated method based on hyperedge sampling to overcome scalability issues of exact algorithms. We have proved that such an approximated algorithm allows for a huge gain in the running time at only a little expense on the accuracy of the results. Moreover, our sampling algorithm allows for the analysis of larger motifs, which were not computationally feasible with exact methods. We believe that faster algorithms for higher-order motif analysis, such as our proposed sampling algorithm, can pave the way for exciting applications. In this direction, important aspects for future work are the development of sampling algorithms with strong approximation guarantees and the investigation of more efficient sampling strategies for the different categories of patterns of higher-order interactions.

\section*{Declarations}

\paragraph{Conflict of interest} The authors declare no conflict of interest.

\paragraph{Acknowledgements} 
F.B. acknowledges support from the Air Force Office of Scientific Research under award number FA8655-22-1-7025. A.M. acknowledges support from the European Union through Horizon Europe CLOUDSTARS project (101086248).

\bibliography{sn-bibliography}% common bib file


\begin{appendices}
\section{Parameters search}
\label{appendix}
\begin{figure}[ht]
    \centering
    \includegraphics[width=\linewidth]{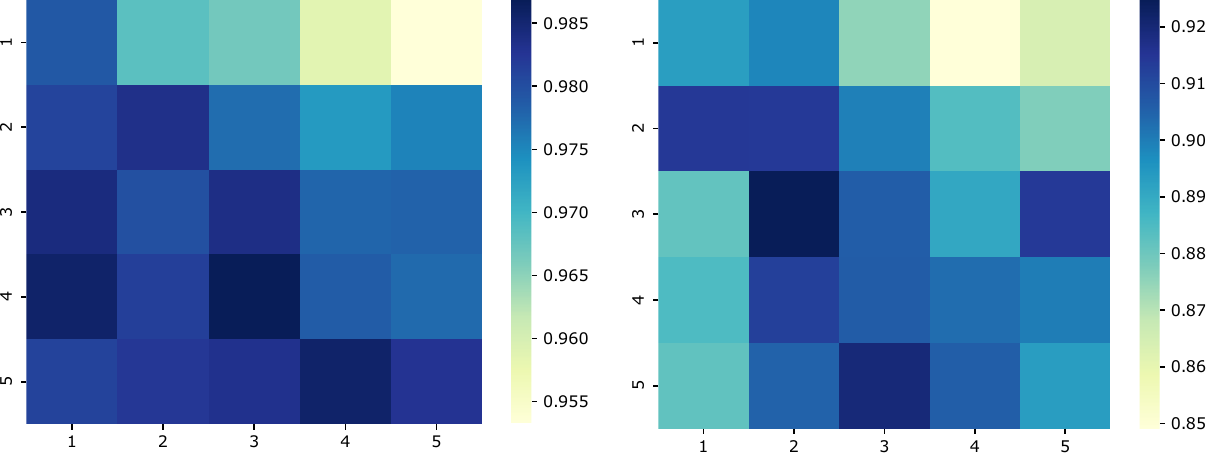}
    \caption{We parametrize the number of samples of hyperedges of size $3$ and $4$ with respect to the number of samples of hyperedges of size $2$ and search the values for which the correlation between the exact motif profile and the estimated one is maximized. The $x$-axis parametrizes the number of samples of hyperedges of size $4$. The $y$-axis parametrizes the number of samples of hyperedges of size $3$. Light squares exhibit lower levels of correlation, while dark squares show higher levels. On the left, we show the matrix for the high school dataset. On the right, is the matrix for the history dataset. We get the best parameters by averaging the two matrices.}
    \label{fig:parameters}
\end{figure}
\end{appendices}
Our approximated algorithm requires different parameters. The parameter $S$ controls the number of samples of hyperedges to be performed to estimate the count of the patterns of sub-hypergraphs. Without a careful design, directly sampling hyperedges from the hypergraphs leads to unreliable results. In fact, the distribution of the size of the hyperedges in a real-world hypergraph is not uniform, leading the algorithm to often sample hyperedges of size $2$, and rarely, for example, hyperedges of size $4$. This would result in poor estimations of higher-order motifs involving a group interaction of size $3$ or $4$.
To address this issue, we stratify our sampling process, allocating specific sample budgets to hyperedges of different sizes. This ensures a balanced representation of hyperedges across all sizes. 
Let $S_k$ be the number of samples assigned to hyperedges of size $k$. We fix the sum of $S_k$ for every $k$ to be equal to $S$. We estimate empirically good values for the parameters $S_k$, exhaustively searching among different combinations of values and selecting those that maximize a defined quality function (Pearson's correlation $\rho$ between the exact higher-order motif profile and the estimated one). We perform the analyses on two datasets, one for each macro-domain: \texttt{high school} and \texttt{history}. We consider motifs of order $4$, therefore we need to estimate $S_2$, $S_3$ and $S_4$, namely, respectively the number of samples from the hyperedges of order $2$, $3$ and $4$. Given that $S_2 + S_3 + S_4 = S$, one can fix $S_2$, parametrize $S_3$ and $S_4$ to be multiple of $S_2$, and perform exhaustive search. We show the results in~\Cref{fig:parameters}. Averaging the results of the two matrices, we get that our quality measure is maximized when $S_3 = 3S_2$ and $S_4 = 2S_2$. We use these parameters in our experiments.

\end{document}